\documentclass{llncs}

\usepackage{amsfonts,amssymb}
\def\FF2{\mbox{$\mathbb F_2$}}

\begin{document}

\title{On a Polytime Factorization Algorithm for Multilinear Polynomials over \FF2%
\thanks{This work is supported by the Ministry of Science and Education of the Russian Federation under the 5--100 Excellence Program and the grant of Russian Foundation for Basic Research No. 17--51--45125.}}

\author{Pavel Emelyanov\inst{1,2}
  \and  Denis Ponomaryov\inst{1,2}}

\institute{A.P. Ershov Institute of Informatics Systems, Lavrentiev av. 6, 630090, Novosibirsk, Russia \and Novosibirsk State University, Pirogov st. 1, 630090, Novosibirsk, Russia \newline
\email{\{emelyanov,ponom\}@iis.nsk.su}}

\maketitle

\begin{abstract}
In 2010, A. Shpilka and I. Volkovich established a prominent result on the equivalence of polynomial factorization and identity testing. It follows from their result that a multilinear polynomial over the finite field of order 2 can be factored in time cubic in the size of the polynomial given as a string. Later, we have rediscovered this result and provided a simple factorization algorithm based on  computations over derivatives of multilinear polynomials. The algorithm has been applied to solve problems of compact representation of various combinatorial structures, including Boolean functions and relational data tables. In this paper, we describe an improvement of this factorization algorithm and report on preliminary experimental analysis.

\end{abstract}

\section{Introduction}
Polynomial factorization is a classic algorithmic problem in algebra \cite{vonzurGathenGerhard-2013}, whose importance stems from numerous applications. The computer era has stimulated interest to  polynomial factorization over finite fields. For a long period of time, Theorem 1.4 in \cite{Grigoriev-1984en} (see also \cite[Th.~1.6]{Shparlinski1992}) has been the main source of information on the complexity of this problem: a (densely represented) polynomial $F_{p^r}(x_1,\ldots ,x_m)$ of the total degree $n>1$ over all its variables can be factored in time that is polynomial in $n^m$, $r$, and $p$. In addition, practical probabilistic factorization algorithms have been known.

In 2010, A. Shpilka and I. Volkovich \cite{ShpilkaVolkovich-2010icalp} established a  connection between polynomial factorization and polynomial identity testing. The result has been formulated in terms of the arithmetic circuit representation of polynomials. It follows from these results that a multilinear polynomial over \FF2 (the finite field of the order 2)  can be factored in the time that is cubic in the size of the polynomial given as a symbol sequence.

Multilinear polynomials over \FF2 are well known in the scope of mathematical logic (as Zhegalkine polynomials \cite{Zhegalkin-1928} or Algebraic Normal Form) and in circuit synthesis (Canonical Reed--Muller Form \cite{Muller-1954}). Factorization of multilinear polynomials is a particular case of decomposition (so--called conjunctive or AND--decomposition) of logic formulas and Boolean functions. By the idempotence law in the algebra of logic, multilinearity (all variables occur in degree 1) is a natural property of these polynomials, which makes the factors have disjoint sets of variables $F(X,Y)=F_1(X)F_2(Y)$, $X\cap Y=\varnothing$. In practice, this property allows for obtaining a factorization algorithm by variable partitioning (see below).

Among other application domains, such as game and graph theory, the most attention has been given to decomposition of Boolean functions in logic circuit synthesis, which is related to the algorithmic complexity and practical issues of electronic circuits implementation, their size, time delay, and power consumption (see \cite{PerkowskiGrygiel-95psudee,KhatriGulati-2011atlsoa}, for example). One may note the renewed interest in this topic, which is due to the novel technological achievements in circuit design.

The logic interpretation of multilinear polynomials over \FF2 admits another notion of factorization, which is commonly called Boolean factorization (finding Boolean divisors). For example, there are Boolean polynomials, which have decomposition components sharing some common variables. Their product/con\-junc\-tion does not produce original polynomials in the algebraic sense but it gives the same functions/formulas in the logic sense. In general, logic--based approaches to decomposition are more powerful than algebraic ones: a Boolean function can be decomposable logically, but not algebraically \cite[Ch. 4]{KhatriGulati-2011atlsoa}.

In 2013, the authors have rediscovered the result of A. Shpilka and I. Volkovich under simpler settings and in a simpler way \cite{EmelyanovPonomaryov-2014psi,EmelyanovPonomaryov-2015pcs}. A straightforward treatment of sparsely represented multilinear polynomials over \FF2 gave the same worst--case cubic complexity of the factorization algorithm. Namely, the authors provided two factorization algorithms based, respectively, on computing the greatest common divisor (GCD) and  formal derivatives (FD) for polynomials obtained from the input one.

The algorithms have been used to obtain a solution to the following problems of compact representation of different combinatorial structures (below we provide examples, which intuitively explain their relation to the factorization problem).
\begin{itemize}
\item Conjunctive disjoint decomposition of monotone Boolean functions given in DNF \cite{EmelyanovPonomaryov-2014psi,EmelyanovPonomaryov-2015pcs}. For example, the following DNF
\begin{equation}
\varphi = (x\wedge u) \ \vee\ (x\wedge v) \ \vee\ (y\wedge u) \ \vee\ (y\wedge v) \vee (x\wedge u \wedge v)
\end{equation}
is equivalent to
\begin{equation}
\psi = (x\wedge u) \ \vee\ (x\wedge v) \ \vee\ (y\wedge u) \ \vee\ (y\wedge v)
\end{equation}
since the last term in $\varphi$ is redundant, and we have
\begin{equation}
\psi\equiv (x\vee y) \wedge (u\vee v)
\end{equation}
and the decomposition components $x\vee y$ and $u \vee v$ can be recovered from the factors of the polynomial
\begin{equation}
F_\psi = xu + xv + yu+ yv = (x+y)\cdot (u+v)
\end{equation}
constructed for $\psi$. \smallskip

\item Conjunctive disjoint decomposition of Boolean functions given in full DNF \cite{EmelyanovPonomaryov-2014psi,EmelyanovPonomaryov-2015pcs}. For example, the following full DNF

\medskip

$\begin{array}{ll}
\varphi =  (x\wedge \neg y \wedge u\wedge \neg v) & \vee (x\wedge \neg y \wedge \neg u \wedge v)\vee \\
& \vee (\neg x \wedge y\wedge u\wedge \neg v) \vee (\neg x \wedge y \wedge \neg u \wedge v)
\end{array}$

is equivalent to
\begin{equation}
(x\wedge \neg y) \vee (\neg x \wedge y) \bigwedge (u\wedge \neg v) \vee (\neg u \wedge v)
\end{equation}
and the decomposition components of $\varphi$ can be recovered from the factors of the polynomial
\begin{equation}
F_\varphi = x\bar{y}u\bar{v} + x\bar{y}\bar{u}v + \bar{x}yu\bar{v}+ \bar{x}y\bar{u}v = (x\bar{y}+\bar{x}y) \cdot (u\bar{v}+\bar{u}v)
\end{equation}
constructed for $\varphi$.  \smallskip

\item Non-disjoint conjunctive decomposition of multilinear polynomials over \FF2, in which components can have common variables from a given set. In \cite{Emelyanov-2016caldam}, a fixed--parameter polytime decomposition algorithm has been proposed, for the parameter being the  number of the shared variables between components.  \smallskip

\item Cartesian decomposition of data tables (i.~e., finding tables such that their unordered Cartesian product gives the source table) \cite{EmelyanovPonomaryov-2017ssdse,Emelyanov-2018iccs} and generalizations thereof for the case of a non-empty subset of shared attributes between the tables.  For example, the following table has a decomposition of the form:

{\footnotesize
\begin{center}
\begin{tabular}{cccccc}

\begin{tabular}{|c|c|c|c|c|}
\hline
B & E & D & A & C \\
\hline\hline
z & q & u & x & y \\
\hline
y & q & u & x & y \\
\hline
y & r & v & x & z \\
\hline
z & r & v & x & z \\
\hline
y & p & u & x & x \\
\hline
z & p & u & x & x \\
\hline
\end{tabular}
& $~=~$ &
\begin{tabular}{|c|c|}
\hline
A & B \\
\hline\hline
x & y \\
\hline
x & z \\
\hline
\end{tabular}
& $\mbox{\large$\times$}$ &
\begin{tabular}{|c|c|c|}
\hline
C & D & E \\
\hline\hline
x & u & p \\
\hline
y & u & q \\
\hline
z & v & r \\
\hline
\end{tabular}
&
\hspace*{3mm}
\end{tabular}
\end{center}
}

which can be obtained from the factors of the polynomial

\medskip

$\begin{array}{ll}
z_B\cdot{}q\cdot{}u\cdot{}x_A\cdot{}y_C + & y_B\cdot{}q\cdot{}u\cdot{}x_A\cdot{}y_C + \\ y_B\cdot{}r\cdot{}v\cdot{}x_A\cdot{}z_C + & z_B\cdot{}r\cdot{}v\cdot{}x_A\cdot{}z_C\, + \\
y_B\cdot{}p\cdot{}u\cdot{}x_A\cdot{}x_C + & z_B\cdot{}p\cdot{}u\cdot{}x_A\cdot{}x_C = \\
& =  (x_A\cdot{} y_B+ x_A\cdot{}z_B) \cdot (q\cdot{}u\cdot{}y_C +r\cdot{}v\cdot{}z_C +p\cdot{}u\cdot{}x_C)
\end{array}$

\medskip

constructed for the table's content. \smallskip

In terms of SQL, Cartesian decomposition means reversing the first operator and the second operator represents some feasible generalization of the problem:

\medskip

\begin{tabular}{ccl}
\tt T1 CROSS JOIN T2 & \hspace*{22mm}  & \tt SELECT T1.*, T2.* EXCEPT(Attr2)\\
                     &                 & \tt \hspace*{3.7mm}FROM T1 INNER JOIN T2\\
                     &                 & \tt \hspace*{7.2mm}ON T1.Attr1 = T2.Attr2
\end{tabular}

\medskip

where {\tt EXCEPT(list)} is an informal extension of SQL used to exclude {\tt list} from the resulting attributes. This approach can be applied to other table--based structures (for example, decision tables or datasets appearing in the K\&DM domain, as well as the truth tables of Boolean functions).
\end{itemize}

Shpilka and Volkovich did not address the problems of practical implementations of the factorization algorithm. However, the applications above require a factorization algorithm to be efficient enough on large polynomials. In this paper, we propose an improvement of the factorization algorithm from \cite{EmelyanovPonomaryov-2017ssdse,Emelyanov-2018iccs}, which potentially allows for working with larger inputs. An implementation of this version of the algorithm in Maple 17 outperforms the native Maple's {\tt Factor(poly) mod 2} factorization, which in our experiments failed to terminate on input polynomials having $10^3$ variables and $10^5$ monomials.

\section{Definitions and Notations}

A polynomial $F\in\FF2[x_1,\ldots,x_n]$ is called \emph{factorable} if $F=F_1\cdot \ldots \cdot F_k$, where $k\geq 2$ and $F_1, \ldots, F_k$ are non-constant polynomials. The polynomials $F_1, \ldots, F_k$ are called \emph{factors} of $F$. It is important to realize that since we consider multilinear polynomials (every variable can occur only in the power of $\leq 1$), the factors are polynomials \emph{over disjoint sets of variables}. In the following sections, we assume that the polynomial $F$ does not have \emph{trivial divisors}, i.e., neither $x$, nor $x+1$ divides $F$. Clearly, trivial divisors can easily be recognized.

For a polynomial $F$, a variable $x$ from the set of variables $Var(F)$ of $F$, and a value $a\in\{0,1\}$, we denote by $F_{x=a}$ the polynomial obtained from $F$ by substituting $x$ with $a$. For multilinear polynomials over \FF2, we define a \emph{formal derivative} as $\frac{\partial{F}}{\partial{x}}=F_{x=0}+F_{x=1}$, but for non-linear ones, we use the definition of a ``standard'' formal derivative for polynomials. Given a variable $z$, we write $z | F$ if $z$ divides $F$, i.e., $z$ is present in every monomial of $F$ (note that this is equivalent to the condition $\frac{\partial{F}}{\partial{z}}=F_{z=1}$).

Given a set of variables $\Sigma$ and a monomial $m$, the \emph{projection} of $m$ onto $\Sigma$ is $1$ if $m$ does not contain any variable from $\Sigma$, or is equal to the monomial obtained from $m$ by removing all the variables not contained in $\Sigma$, otherwise. The \emph{projection} of a polynomial $F$ onto $\Sigma$, denoted as $F|_\Sigma$, is the polynomial obtained as sum of monomials from the set $S$, where $S$ is the set of the monomials of $F$ projected onto $\Sigma$.

$|F|$ is the \emph{length} of the polynomial $F$ given as a symbol sequence, i.e., if the polynomial over $n$ variables has $M$ monomials of lengths $m_1,\ldots,m_M$ then $|F|=\sum_{i=1}^{M}m_i=O(nM)$.

We note that the correctness proofs for the algorithms presented below can be found in \cite{EmelyanovPonomaryov-2014psi,EmelyanovPonomaryov-2015pcs}.

\section{GCD--Algorithm}

Conceptually, this algorithm is the simplest one. It outputs factors of an input polynomial whenever they exist.
\newpage
\begin{enumerate}
\item{\tt Take an arbitrary variable $x$ from $Var(F)$}

\item{\tt $G:=\gcd(F_{x=0},\frac{\partial{F}}{\partial{x}})$}

\item{\tt If $G=1$ then stop}

\item{\tt Output factor $\frac{F}{G}$}

\item{\tt $F:=G$}

\item{\tt Go to 1}
\end{enumerate}
\noindent Here the complexity of factorization is hidden in the algorithm for finding the greatest common divisor of polynomials.

Computing GCD is known as a classic algorithmic problem in algebra \cite{vonzurGathenGerhard-2013}, which involves computational difficulties. For example, if the  field is not too rich (\FF2 is an example) then intermediate values vanish quite often, which essentially affects the computation performance. In \cite{deKleineMonaganWittkopf-2005}, A. Wittkopf et al. developed the LINZIP algorithm for the GCD--problem. Its complexity is $O(|F|^3)$, i.~e., the complexity of the GCD--algorithm is asymptotically the same as for Shpilka and Volkovich's result for the case of multilinear polynomials (given as strings).

\section{FD--Algorithm}

In the following, we assume that the input polynomial $F$ contains at least two variables. The basic idea of FD--Algorithm is to partition a variable set into two sets with respect to a selected variable:
\vspace{-1.5mm}
\begin{itemize}
\item the first set $\Sigma_{same}$ contains the selected variable and corresponds to
an irreducible polynomial;

\item the second set $\Sigma_{other}$ corresponds to the second polynomial that can admit further factorization.
\end{itemize}
\vspace{-1.5mm}
As soon as $\Sigma_{same}$ and $\Sigma_{other}$ are computed (and $\Sigma_{other}\neq\varnothing$), the corresponding factors can be easily obtained as projections of the input polynomial onto these sets.
\begin{enumerate}
\item{\tt Take an arbitrary variable $x$ occurring in $F$}

\item{\tt Let $\Sigma_{same}:=\{x\}, \Sigma_{other}:=\varnothing, F_{same}:=0, F_{other}:=0$}

\item\label{Prod_Computing}
     {\tt Compute $G:=F_{x=0}\cdot \frac{\partial{F}}{\partial{x}}$}

\item\label{FD_Checking}
     {\tt For each variable $y\in Var(F)\setminus\{x\}$: \\
          \hspace*{3mm}If $~\frac{\partial{G}}{\partial{y}}=0~$
                          then $~\Sigma_{other}:=\Sigma_{other}\cup\{y\}~$ \\
          \hspace*{25.5mm}else $~\Sigma_{same}:=\Sigma_{same}\,\cup\{y\}$}

\item{\tt If $\Sigma_{other}\!=\!\varnothing$ then report $^{\prime\prime}F$ is non-factorable$\,^{\prime\prime}$ and stop}

\item{\tt Return polynomials $F_{same}$ and $F_{other}$ obtained as projections onto $\Sigma_{same}$ and $\Sigma_{other}$, respectively}

\end{enumerate}
\noindent The factors $F_{same}$ and $F_{other}$ have the property mentioned above and hence,  the algorithm can be applied to obtain factors for $F_{other}$.

\medskip

Note that FD--algorithm takes $O(|F|^2)$ steps to compute the polynomial $G=F_{x=0}\cdot \frac{\partial{F}}{\partial{x}}$ and $O(|G|)$ time to test whether the derivative $\frac{\partial{G}}{\partial{y}}$ equals zero. As we have to verify this for every variable $y\neq x$, we have a procedure that computes a variable partition in $O(|F|^3)$ steps. The algorithm allows for a straightforward parallelization on the selected variable $y$:
the loop over the variable $y$ (selected in line \ref{FD_Checking}) can be performed in parallel for all the variables.


One can readily see that the complexity of factorization is hidden in the computation of the product $G$ of two polynomials and testing whether a derivative of this product is equal to zero. In the worst case, the length of $G=F_{x=0}\cdot \frac{\partial{F}}{\partial{x}}$ equals $\Omega(|F|^2)$, which makes computing this product expensive for large input polynomials. In the next section, we describe a modification of the FD-algorithm, which implements the test above in a more efficient recursive fashion, without the need to compute the product of polynomials explicitly.

\section{Modification of FD--Algorithm}

Assume the polynomials $A=\frac{\partial{F}}{\partial{x}}$ and $B=F_{x=0}$ are computed. By taking a derivative of $A\cdot B$ on $y$ (a variable different from $x$) we have $D=\frac{\partial{F_{x=0}}}{\partial{y}}$ and $C=\frac{\partial^2{F}}{\partial{x}\partial{y}}$. We need to test whether $AD+BC=0$, or equivalently, $AD=BC$. The main idea is to reduce this test to four tests involving polynomials of smaller sizes. Proceeding recursively in this way, we obtain smaller, or even constant, polynomials for which identity testing is simpler. Yet again, the polynomial identity testing demonstrates its importance, as Shpilka and Volkovich have readily established.

\smallskip

Steps \ref{Prod_Computing}--\ref{FD_Checking} of FD--algorithm are modified as follows:

\vskip\baselineskip
\begin{minipage}{\textwidth}
{\tt
Let $A=\frac{\partial{F}}{\partial{x}}$, $B=F_{x=0}$\\
For each variable $y\in Var(F)\setminus\{x\}$: \\
\hspace*{3mm}Let $D=\frac{\partial{B}}{\partial{y}}$, $C=\frac{\partial{A}}{\partial{y}}$\\
\hspace*{3mm}If IsEqual($A$,$D$,$B$,$C$) then $~\Sigma_{other}:=\Sigma_{other}\cup\{y\}$, \\
\hspace*{42mm}                            else $~\Sigma_{same}:=\Sigma_{same}\cup\{y\}$
}
\end{minipage}
\vskip\baselineskip

\noindent where (all the above mentioned variables are chosen from the set of variables of the corresponding polynomials).

\vskip\baselineskip
\noindent{\tt Define IsEqual($A$,$D$,$B$,$C$) returning Boolean}
\begin{enumerate}
\item\label{BegTGroup1}%
     {\tt If $A=0$ or $D=0$ then return ($B=0$ or $C=0$)}

\item\label{EndTGroup1}%
     {\tt If $B=0$ or $C=0$ then return FALSE}

\medskip

\item\label{BegFor}%
     {\tt For all variables $z$ occurring in at least one of $A,B,C,D$ :}
\item \hspace{2mm} {\tt If ($z | A$ or $z | D$) xor ($z | B$ or $z | C$) then return FALSE}
\item \hspace{2mm} {\tt  Replace every $X\in\{A,B,C,D\}$ with $X:=\frac{\partial{X}}{\partial{z}}$, provided $z|X$ }
\medskip

\item\label{BegTGroup2}%
     {\tt If $A=1$ and $D=1$ then return ($B=1$ and $C=1$)}

\item{\tt If $B=1$ and $C=1$ then return FALSE}

\item{\tt If $A=1$ and $B=1$ then return ($D=C$)}

\item\label{EndTGroup2}%
     {\tt If $D=1$ and $C=1$ then return ($A=B$)}

\medskip



\item\label{PickVar}%
     {\tt Pick a variable $z$} \smallskip

\item\label{FirstProcCall}%
{\tt If not IsEqual($A_{z=0}$,$D_{z=0}$,$B_{z=0}$,$C_{z=0}$) then return FALSE} \smallskip

\item\label{SecondProcCall}%
{\tt If not IsEqual($\,\,\,\,\,\,\frac{\partial{A}}{\partial{z}}$,\,\,\,\,\,$\frac{\partial{D}}{\partial{z}}$,\,\,\,\,\,$\frac{\partial{B}}{\partial{z}}$,\,\,\,\,\,$\frac{\partial{C}}{\partial{z}}$) then return FALSE} \smallskip

\item\label{ThirdProcCall}%
{\tt If \hspace*{7.3mm}IsEqual($\,\,\,\,\,\,\frac{\partial{A}}{\partial{z}}$,$\,B_{z=0}$,$A_{z=0}$,$\,\,\,\,\frac{\partial{B}}{\partial{z}}$) then return TRUE} \smallskip

\item\label{FourthProcCall}%
{\tt Return IsEqual($\,\,\,\,\,\,\frac{\partial{A}}{\partial{z}}$,$\,C_{z=0}$,$A_{z=0}$,\,\,\,\,$\frac{\partial{C}}{\partial{z}}$)}

\end{enumerate}
\noindent{\tt End Definition}
\vskip\baselineskip

Several comments on {\tt IsEqual} are in order:

\begin{itemize}
\item Lines \ref{BegTGroup1}--\ref{EndTGroup2} implement processing of trivial cases, when the condition $AD = BC$ can easily be verified without recursion. For example, when line 2 is executed, it is already known that neither $A$, nor $D$ equals zero and hence, $AD$ can not be equal to $BC$. Similar tests are implemented in lines 6--9. \smallskip

\item At line 5 it is known that $z$ divides both, $AD$ and $BC$ and thus, the problem $AD = BC$ can be reduced to the polynomials obtained by eliminating $z$. \smallskip

\item Finally, lines \ref{FirstProcCall}--\ref{FourthProcCall} implement recursive calls to {\tt IsEqual}. Observe that the parameter polynomials are obtained from the original ones by evaluating a variable $z$ to zero or by computing a derivative. The both operations yield polynomials of a smaller size than the original ones and can give constant polynomials in the limit. To determine the parameters of {\tt IsEqual} we resort to a trick that transforms one identity into two smaller ones. This transformation uses a multiplier, which is not unique. Namely, we can select 16 variants among 28 possible ones (see comments in Section \ref{AllParameters} below) and this gives 16 variants of lines \ref{ThirdProcCall}--\ref{FourthProcCall}. 
\end{itemize}

\subsection{Complete List of Possible Parameters}\label{AllParameters}

If $A$, $D$, $B$, $C$ are the parameters of {\tt IsEqual}, we denote for a $Q\in\{A,D,B,C\}$ the derivative on a variable $z$ and evaluation $z=0$ as $Q_1$ and $Q_2$, respectively.
\[
AD=BC \hspace*{3mm}\mbox{iff}\hspace*{3mm}  (A_1z+A_2)(D_1z+D_2)=(B_1z+B_2)(C_1z+C_2),
\]
\[
A_1D_1z^2+(A_1D_2+A_2D_1)z+A_2D_2=B_1C_1z^2+(B_1C_2+B_2C_1)z+B_2C_2.
\]
The equality holds iff the corresponding coefficients are equal:
\[
\left\{
\begin{array}{rclc}
A_1D_1 &=& B_1C_1               &  (1)\\
A_2D_2 &=& B_2C_2               &  (2)\\
A_1D_2+A_2D_1 &=& B_1C_2+B_2C_1 &  (3)\\
\end{array}
\right.
\]
If at least one of the identities (1), (2) does not hold then $AD\neq BC$. Otherwise, we can use these identities to verify (3) in the following way.

By the rule of choosing $z$, we can assume $A_1,A_2\neq 0$. Multiplying both sides of $(3)$ by $A_1A_2$ gives
\[
A_1^2A_2D_2+A_1A_2^2D_1 = A_1A_2B_1C_2+A_1A_2B_2C_1.
\]
Next, by using the identities (1) and (2),
\[
A_1^2B_2C_2+A_1A_2B_2C_1 = A_2^2B_1C_1+A_1A_2B_1C_2,
\]
\vskip-1.3\baselineskip
\[
A_1B_2(A_1C_2+A_2C_1) = A_2B_1(A_2C_1+A_1C_2).
\]
Hence, it suffices to check $(A_1B_2+A_2B_1)(A_1C_2+A_2C_1) = 0$, i.e., at least one of these factors equals zero. It turns out that we need to test at most 4 polynomial identities, and each of them is smaller than the original identity $AD=BC$. Notice that the multiplier $A_1A_2$ is used to construct the version of {\tt IsEqual} given above.

By the rule of choosing $z$, we can take different multiplier's combinations of the pairs of 8 elements. Only 16 out of 28 pairs are appropriate
\[
\begin{array}{ccc}
A_1 A_2  & \rightarrow &  A_1 C_2 = A_2 C_1,~~ A_1 B_2 = A_2 B_1 \\
A_1 B_2  & \rightarrow &  A_1 D_2 = B_2 C_1,~~ A_1 B_2 = A_2 B_1 \\
A_1 C_2  & \rightarrow &  A_1 D_2 = B_1 C_2,~~ A_1 C_2 = A_2 C_1 \\
A_1 D_2  & \rightarrow &  A_1 D_2 = B_2 C_1,~~ A_1 D_2 = B_1 C_2 \\
A_2 B_1  & \rightarrow &  A_2 D_1 = B_1 C_2,~~ A_1 B_2 = A_2 B_1 \\
A_2 C_1  & \rightarrow &  A_2 D_1 = B_2 C_1,~~ A_1 C_2 = A_2 C_1 \\
A_2 D_1  & \rightarrow &  A_2 D_1 = B_2 C_1,~~ A_2 D_1 = B_1 C_2 \\
B_1 B_2  & \rightarrow &  B_1 D_2 = B_2 D_1,~~ A_1 B_2 = A_2 B_1 \\
B_1 C_2  & \rightarrow &  A_2 D_1 = B_1 C_2,~~ A_1 D_2 = B_1 C_2 \\
B_1 D_2  & \rightarrow &  B_1 D_2 = B_2 D_1,~~ A_1 D_2 = B_1 C_2 \\
B_2 C_1  & \rightarrow &  A_2 D_1 = B_2 C_1,~~ A_1 D_2 = B_2 C_1 \\
B_2 D_1  & \rightarrow &  B_1 D_2 = B_2 D_1,~~ A_2 D_1 = B_2 C_1 \\
C_1 C_2  & \rightarrow &  C_1 D_2 = C_2 D_1,~~ A_1 C_2 = A_2 C_1 \\
C_1 D_2  & \rightarrow &  C_1 D_2 = C_2 D_1,~~ A_1 D_2 = B_2 C_1 \\
C_2 D_1  & \rightarrow &  C_1 D_2 = C_2 D_1,~~ A_2 D_1 = B_1 C_2 \\
D_1 D_2  & \rightarrow &  C_1 D_2 = C_2 D_1,~~ B_1 D_2 = B_2 D_1 \\
\end{array}
\]

\subsection{Analysis of ModFD--Algorithm for Random Polynomials}\label{Sect:Analysis}

We now provide a theoretical analysis of ModFD--algorithm. The complexity estimations we describe here are conservative and, therefore, they give an upper bound greater than $O(|F|^3)$ of the original FD--algorithm. However, the approach presented here could serve as a basis to obtain a more precise upper bound, which would  explain the gain in performance in practice; we report
on a preliminary experimental evaluation  in Section \ref{Sect:Discussion}.

Our estimation is based on

\begin{theorem}[Akra and Bazzi, \cite{AkraBazzi-1998:coa}]\label{Akra-Bazzi-Theorem}
Let the recurrence
\[
T(x)=g(x)+\sum_{i=1}^{k}\lambda_iT(\omega_ix+h_i(x))  ~~\mbox{for}~~ x\geq C
\]
satisfy the following conditions:
\begin{enumerate}
\item $T(x)$ is appropriately defined for $x<C$;
\item $\lambda_i>0$ and $0<\omega_i<1$ are constants for all $i$;
\item $|g(x)|=O\left(x^{c}\right)$; and
\item $|h_i(x)|=O\left(\frac{x}{(\log{x})^2}\right)$  for all $i$.
\end{enumerate}
Then
\[
T(x)=\Theta\left(x^p\left(1+\int_{1}^{x}\frac{g(t)}{t^{p+1}}dt\right)\right),
\]
where $p$ is determined by the characteristic equation $\sum_{i=1}^{k}\lambda_i\omega_i^p=1$.
\end{theorem}

For the complexity estimations, we assume that polynomials are represented by alphabetically sorted lists of bitscales corresponding to indicator vectors for the variables of monomials. Hence, to represent a polynomial $F$ over $n$ variables with $M$ monomials $|F|=nM+cM$ bits are required, where $c$ is a constant overhead to maintain the list structure. This guarantees the linear time complexity for the following operations:
\begin{itemize}
\item computing a derivative with respect to a variable (the derived polynomial also remains sorted);

\item evaluating to zero for a variable with removing the empty bitscale representing the constant 1 if it occurs  (the derived polynomial also remains sorted);

\item identity testing for polynomials derived from the original sorted polynomial by the two previous operations.
\end{itemize}

For {\tt IsEqual} we have

\begin{enumerate}
\item $x=|A|+|B|+|C|+|D|$. By taking into account the employed representation of monomials (the bitscale is not shortened when a variable is removed), we may also assume that $|Q|=|Q_1|+|Q_2|$.

\medskip

\item $\forall i, \lambda_i=1$.

\medskip

\item $\forall i, h_i(x)=0$.

\medskip

\item $g(x)=O(nx)$.
Therefore, the total time for lines \ref{BegTGroup1}--\ref{PickVar} consists of the constant numbers of linear (with respect to the input of {\tt IsEqual}) operations executed at most $n$ times. Apparently, $n$ is quite a conservative assumption, because at a single recursion step, at least one variable is removed from the input set of variables.

\medskip

\item We need to estimate $\omega_1$, $\omega_2$, $\omega_3$, $\omega_4$. \\
Among all the possible choices of the multipliers mentioned in Section \ref{AllParameters}, let us consider those of the form $Q_1Q_2$. They induce two equations that do not contain one of the input parameters of {\tt IsEqual}: $A$, $B$, $C$, $D$ result in the absence of the parts of $D$, $C$, $B$, $A$, respectively,  among the parameters of {\tt IsEqual} in lines \ref{ThirdProcCall} and \ref{FourthProcCall}. Hence, the largest parameter can be excluded by taking an appropriate $Q$; lines \ref{ThirdProcCall}--\ref{FourthProcCall} of ModFD--algorithm are to be rewritten with the help of this observation.

\medskip

Without loss of generality, we may assume that the largest parameter is $D$ and thus, we can take $Q$ equal to $A$. In this case, $\omega_1$, $\omega_2$, $\omega_3$, $\omega_4$ represent the relative lengths of the parameters $|A_1|+|B_1|+|C_1|+|D_1|$, $|A_2|+|B_2|+|C_2|+|D_2|$, $|A|+|B|$, $|A|+|C|$ for the recursive calls to {\tt IsEqual} with respect to $|A|+|B|+|C|+|D|$.

\medskip

Since $|A|, |B|, |C| \leq |D|$, we obtain $|A|+|B|$, $|A|+|C|$, $|B|+|C| \leq 2|D|$. Then the lengths $|A|+|B|$ and $|A|+|C|$,
respectively, can be estimated in the following way:

\[
|A|+|B|=x-|C|-|D|\leq x-0-\frac{|A|+|B|}{2},
\]
hence, $|A|+|B|$, $|A|+|C|\leq\frac23$.

\medskip

Let $F$ be a multilinear polynomial over $n$ variables with $M$ monomials such that no variable divides $F$. A random polynomial consists of monomials randomly chosen from the set of all monomials over $n$ variables. Variables appear in monomials independently.
For each variable $x$ from {\tt var}$(F)$, we can consider the following quantity $\mu_x=|F_x^\prime|$ (i.e. the part of monomials containing this variable). We want to estimate the probability that among $\mu_x$ there exist at least one, which is (approximately) equal to $\frac{M}{2}$. Hence
\[
\begin{array}{rcl}
\mathbb{P}\left[\mbox{there exists $x$ such that $\mu_x$ is a median}\right]
& = &
1-\mathbb{P}\left[\bigwedge_x \mu_x ~\mbox{is not median}\right] \\
& = &
1-\mathbb{P}\left[\mu_1 ~ \mbox{is not median}\right]^n \\
& = &
1-\left(1-\mathbb{P}\left[\mu_1 ~ \mbox{is a median}\right]\right)^n \\
& = &
1-\left(1-\frac12\right)^n \\
& = &
1-\frac{1}{2^n}
\end{array}
\]
Thus, with a high probability one can pick from a large polynomial (in our case, from $D$) a variable such that $|D_1|\approx |D_2|$.

\medskip

Let us consider the following multicriteria linear program
\[
\begin{array}{llll}
\mbox{maximize} &
\left\{
\begin{array}{c}
a_1+b_1+c_1+d_1\\
a_2+b_2+c_2+d_2\\
a+b\\
a+c
\end{array}
\right\} &
\mbox{subject to} &
\left\{
\begin{array}{c}
a+b+c+d=1\\
d_1=d_2\\
a\leq d, ~b\leq d, ~c\leq d \\
a+b\leq\frac23 \\
a+c\leq\frac23 \\
\mbox{all nonnegative}
\end{array}
\right.
\end{array}
\]
where the small letters stand for the sizes of the polynomials denoted by capital letters, e.g., $a$ stands for $|A|$.
Since the objective functions and  constraints are linear and the optimization domain is bounded, we can enumerate all the extreme points of the problem and select those points that give the maximum solution of the characteristic equation of Theorem \ref{Akra-Bazzi-Theorem}. By taking into account the symmetries between the first and the second objective functions and between the third and fourth ones, we obtain that
\[
\omega_1=\frac34, ~~\omega_2=\frac14, ~~\omega_3=\frac12, ~~\omega_4=\frac12 ~~~~~~~(*)
\]

\end{enumerate}

Hence, the characteristic equation is

\[
\left(\frac34\right)^p+\left(\frac14\right)^p+\left(\frac12\right)^p+\left(\frac12\right)^p=1.
\]
\noindent Its unique real solution is $p\approx2.226552$.
Finally, the total time for the ModFD--algorithm obtained this way is
\[
T=O(n^2|F|^{2.226552}).
\]

\section{Preliminary Experiments and Discussion}\label{Sect:Discussion}

For a computational evaluation of the developed factorization algorithms, we used Maple 17 for Windows run on 3.0 GHz PC with 8 GB RAM. The factorization algorithm implemented in Maple {\tt Factor(poly) mod 2} can process multilinear polynomials over \FF2 with hundreds of variables and several thousands of monomials in several hours. But many attempts of factorization of polynomials with $10^3$ variables and $10^5$ monomials were terminated by the time limit of roughly one week of execution. In general, a disadvantage of all Maple implementations is that they are memory consuming. For example, the algorithm that requires computing products of polynomials fails to work even for rather small examples (about $10^2$ variables and $10^3$ monomials). Although GCD--algorithm is conceptually simple, it involves computing the greatest common divisor for polynomials over the ``poor'' finite field \FF2. A practical implementation of LINZIP is not that simple. An older version of Maple reports on some inputs that "LINZIP/not implemented yet". We did not observe this issue in Maple 17. It would be important to conduct an extensive comparison of the performance of GCD-- and FD--algorithm implemented under similar conditions. The factorization algorithm (FD--based) for sparsely represented multilinear polynomials over \FF2 demonstrates reasonable performance. BDD/ZDD can be considered as some kind of the black--box representation. We are going to implement factorization based on this representation and to compare these approaches.

A careful study of the solution \mbox{$(*)$} given at the end of Section \ref{Sect:Analysis} shows that it describes the case when $|A|\approx|D|\approx\frac{x}{2}$ and $|B|\approx|C|\approx0$. This means that at the next steps the maximal parameter is $A$: $|A|\approx\frac{x}{2}$, while the remaining parameters are smaller. Thus, one can see that the lengths of the inputs to the recursive calls of {\tt IsEqual} are reduced at least twice in at most two levels of the recursion. This allows for obtaining a more precise complexity bound, which will be further studied.

Yet another property is quite important for the performance of the algorithm. Evaluating the predicate {\tt IsEqual} for the variables from the same factor requires significantly less time compared with evaluation for other variables. For polynomials with 50 variables and 100 monomials in the both components, the speed-up achieves 10--15 times. The reason is evident and it again confirms the importance of (Zero) Polynomial Identity Testing, as shown by Shplika and Volkovich. Testing that the polynomial $AD+BC$ is not zero requires less reduction steps in contrast with the case when it does equal zero. The latter requires reduction to the constant polynomials. Therefore, we used the following approach: if the polynomials $A$, $D$, $B$, $C$ are ``small'' enough then the polynomial $AD+BC$ was checked to be zero directly via multiplication. For the polynomials with the above mentioned properties, this allows to save about 3--5\% of the execution time. The first practical conclusion is that in general, the algorithm works faster for non--factorable polynomials than for factorable ones. The second is that we need to investigate new methods to detect variables from the ``opposite'' component (factor). Below we give an idea of a possible approach.

It is useful to detect cases of irreducibility before launching the factorization procedure. Using simple necessary conditions for irreducibility, as well as testing simple cases of variable classification for variable partition algorithms, can substantially improve performance. Let $F$ be a multilinear polynomial over $n$ variables with $M$ monomials such that no variable divides $F$. For each variable $x$, recall that the value $\mu_x$ corresponding to the number of monomials containing $x$, i.~e. the number of monomials in $\frac{\partial{F}}{\partial{x}}$. Then a necessary condition for $F$ to be factorable is
\[
\forall x~~ \gcd\left(\mu_x,~M\right)>1.
\]
In addition, we have deduced several properties, which are based on analyzing occurrences of pairs of variables in the given polynomial (for example, if there is no monomial, in which two variables occur simultaneously then these variables can not belong
 to different factors). Of course, the practical usability of these properties depends on how easily they can be tested.

Finally, we note an important generalization of the factorization problem, which calls for efficient implementations of the factorization algorithm. To achieve a deeper optimization of logic circuits we asked in \cite{EmelyanovPonomaryov-2014psi,EmelyanovPonomaryov-2015pcs} how to find a representation of a polynomial in the form $F(X,Y)=G(X)H(Y)+D(X,Y)$, where a ``relatively small'' defect'' $D(X,Y)$ extends or shrinks the pure disjoint factors.  Yet another problem is to find a representation of the polynomial in the form
\[
F(X,Y)=\sum_{k}G_k(X)H_k(Y), ~~~~X\cap Y=\varnothing,
\]
i.~e., a complete decomposition without any ``defect'', which (along with the previous one) has quite interesting applications in the knowledge and data mining domain. Clearly, such decompositions (for example, the trivial one, where each monomial is treated separately) always exist, but not all of them are meaningful from the K\&DM point of view. For example, one might want to put a restriction on the size of the ``factorable part'' of the input polynomial (e.g., by requiring the size to be maximal), which opens a perspective into a variety of optimization problems. Formulating additional constraints targeting factorization is an interesting research topic. One immediately finds a variety of the known computationally hard problems in this direction and it is yet to be realized how the computer algebra and theory of algorithms can mutually benefit from each other along this way.

\end{document}